\documentclass[aps,prb,twocolumn,longbibliography]{revtex4-2}

\usepackage{graphicx}
\usepackage{amsmath,amssymb}
\usepackage{bm}
\usepackage[version=4]{mhchem} 
\usepackage{siunitx}
\usepackage{braket}
\usepackage{hyperref}
\hypersetup{colorlinks=true,linkcolor=blue,citecolor=blue,urlcolor=blue}
\usepackage{microtype}
\usepackage{subfig}
\usepackage{ragged2e} 
\usepackage{multirow}
\makeatletter
\long\def\@makecaption#1#2{%
  \vskip\abovecaptionskip
  \parbox{\linewidth}{\justifying\small #1\quad #2}\par
  \vskip\belowcaptionskip}
\makeatother

\begin{document}
\title{A systematic study of single molecule metallocenes with 4d and 3d transition metal atoms}

\author{Daniela Herrera-Molina}
\author{Kushantha P. K. Withanage}
\author{Jes\'us N. Pedroza-Montero}
\author{Pardeep Kaur}
\author{Mark. R. Pederson}
\author{M. F. Islam}
\affiliation{Department of Physics, University of Texas at El Paso, El Paso, Texas}

\date{\today}

\begin{abstract}

The realization of spin-based devices remains one of the central goals of spintronics research. Single-molecule magnets (SMMs) constitute an important class of nanoscale magnetic systems with significant potential for spintronic applications, where individual molecules can serve as fundamental building blocks of functional devices. In this work, we systematically investigate a family of 4d and 3d transition-metal metallocenes using first-principles density functional theory. Among the seven 4d metallocenes considered, only Mo and Rh metallocenes undergo Jahn–Teller distortions and exhibit uniaxial anisotropy with energy barriers of approximately 20 K. Similarly, among the 3d metallocenes studied in this work, only Cr and Co metallocenes undergo Jahn–Teller distortions and display uniaxial anisotropy, although with smaller barriers below 10 K. All remaining metallocenes exhibit easy-plane anisotropy. We find that the magnetic anisotropy energy does not increase monotonically with the number of d electrons; instead, it is governed primarily by the orbital ordering of the transition-metal d states. Our calculations further show that the Jahn–Teller distortion induces transverse anisotropy, leading to zero-field magnetization tunneling across the energy barrier, with the weakest tunneling rate for Mo metallocene. For the Mo metallocene, the magnetic anisotropy energy increases to approximately 60 K in cationic charge states, although the magnetic anisotropy changes from uniaxial to easy-plane. In this work, we also investigate the influence of ligand size on the structural stability of metallocenes and establish practical guidelines for constructing reliable molecular models for first-principles studies. Finally, we also propose that metallocenes with easy-plane anisotropy could serve as magnetic sensing elements, highlighting their potential beyond memory applications.

\end{abstract}

\maketitle

\section{Introduction}

Magnetic molecules are metal–organic complexes that carry a stable net spin moment, which originates from a single magnetic ion such as a transition metal or a rare earth atom, or from a collection of exchange-coupled magnetic ions\cite{Sessoli2017,Bogani2008}. The great appeal of these materials is that the quantum ground state of these magnetic molecules can represent fundamental bits of information and can be utilized as magnetic memory devices\cite{Christou2000,Mannini2010,Rinehart2011,WANG2021}, single electron transistors\cite{Heersche2006,Henderson2007,Rossier2007,Nossa2013,Chiesa2019,Zhao2025} or qubits for quantum information processing\cite{Wernsdorfer1999,Leuenberger2001,Lehmann2007,Islam2010,Sessoli2015,Ferrando2016,Urtizberea2018,Moreno2018,Gaita2019}. Theoretically, the inherent locality of magnetic moments with fewer magnetic centers allows the investigation of magnetic states with Hamiltonians of finite dimensions, facilitating the exact treatment of quantum magnetic properties. The availability of many different types of magnetic ions, combined with the flexibility of coordination chemistry\cite{Woodruff2013,Kub2025}, offers the possibility to synthesize molecules with an enormous range of geometries. Recent advances in synthesis techniques have enabled the magnetic properties of molecular magnets to be tailored with remarkable precision\cite{Huang2023}. In comparison to other conventional solid-state magnetic materials, molecular magnets hold several advantages, such as low production costs, compatibility with scaling, low power consumption, and, most of all, high tunability through chemical modifications\cite{Friedman2010}.

Since the synthesis of {Mn$_{12}$}-acetate,\cite{Sessoli1993,Gatteschi2003} a wide variety of single-molecule magnets (SMMs) have been investigated, both theoretically and experimentally, to understand the properties that are essential for practical device applications.\cite{Yin2023} For memory applications, one of the key requirements is a large uniaxial magnetic anisotropy barrier arising from the spin--orbit coupling (SOC) of the molecule,\cite{Gatteschi2006,Moreno2021,Leuenberger2001} which stabilizes two bistable ground states that can serve as the binary bits of information.\cite{Pederson2000,Raghunathan2008,Hill2009,Jacobson2015,Cornia2020,Kabir2020,Cheng2021,Alessio2023,Slavensky2025} A high anisotropy barrier is also essential for protecting the stored information against thermal fluctuations. In addition, for SMMs to function effectively in spintronic applications, both the spin coherence time\cite{Schlegel2008,Shiddiq2016,Bayliss2022,Chen2026} and the spin relaxation time\cite{Waldmann2006,Ho2016,Jackson2021,Gu2022} must be sufficiently long.      

In this work, we have considered metallocenes, a class of organometallic complexes, as SMM. These molecules contain a single 3d or 4d transition metal atom or a rare earth atom, which is the source of magnetism in this complex. Among the 3d metallocenes, ferrocene is the most common metallocene. Getahun \emph{et al.} reported superparamagnetic behavior in an aminoferrocene-graphene composite at room-temperature\cite{Getahun2023}. Beyond 3d systems, lanthanide metallocenes such as Terbium (TbCp$_2$) and Dysprosium (DyCp$_2$) were experimentally observed to have very large uniaxial magnetic anisotropy barriers and displayed a slow magnetic relaxation\cite{Gould2019,Ming2025}. In addition to their use as SMMs, metallocenes have been widely used as homogeneous catalyst precursors\cite{Brintzinger1995,Gansaeuer2000}. Their well-defined coordination environment provides uniform single active sites, enabling precise control over polymer growth and leading to narrow molar mass distributions\cite{kaminsky2001}. Furthermore, their properties can be tuned through changes in the metal center and charge state\cite{Resconi2000}. Most studies of metallocenes reported in the literature have focused on the electronic properties of 3d transition-metal elements.

The motivations of the present work are as follows: (i) Despite promising potential for spintronics applications, 4d metallocenes remain largely unexplored. In this work we have provided a comprehensive study of the electronic and magnetic properties of this class of molecules. Although the electronic properties of 3d metallocenes are well studied \cite{Nawa2016}, their magnetic anisptropic properties are much less explored. One of the main objectives of this work is to investigate whether these two series of metallocenes are suitable SMMs for spintronics applications. (ii) Experimentally, metallocenes can be synthesized in different ligand environments. For computational studies, particlualrly within the framework of first-principles density functional theory (DFT), computational cost increases exponentially with the size of the system. It is, therefore, desirable to perform calculations with a model of ligand environment that contains as few as possible ligand atoms, provided that it does not compromise the electronic properties of the metal center, which essentially determines the magnetic anisotropic properties. In this work, we have studied the dependence of the electronic properties of the metal center with different sizes of ligand models.

To address motivation (i), we have performed electronic and magnetic anisotropic properties calculations of 4d metallocenes using DFT, with particular emphasis on the role of d-orbital filling within the same ligand framework considered in Ref.~\cite{Gould2019}. We also examined how the magnetic anisotropy energy (MAE) varies with the molecular charge state for selected metallocenes. To place the results in context, we additionally investigate several 3d metallocenes, providing a direct comparison of magnetic anisotropic behavior across the two transition-metal series. Our work complements previous studies by establishing benchmarks for 4d metal centers (Y--Rh), where the stronger spin--orbit coupling relative to 3d systems enhances magnetic anisotropy while retaining greater chemical accessibility than their 5d counterparts. Although MAE is a key descriptor for the performance of SMMs, it does not fully determine the stability of their magnetic states. Spin-relaxation processes can substantially reduce the lifetime of these states, thereby limiting their usefulness in spintronics applications. To address this issue, we have developed an effective spin model and investigated magnetization tunneling induced by transverse magnetic anisotropy, which provides an important relaxation pathway. Finally, we study the response of uniaxial metallocenes to an external magnetic field, a property that governs magnetization switching and is therefore essential for spintronics device operation. To address motivation (ii), we have constructed three different ligand models of various sizes and investigated their effect on electronic properties and structural stability.  

The organization of the paper is as follows. In section \ref{computation} we discuss the computational methodology adopted in this work. We then present the results of our calculations. Since these 4d metallocenes are not experimentally realized within the same ligand environment, we have first studied their structural stability, which is discussed in subsection \ref{stability}. Furthermore, we discuss the role of the size of the ligand-group in the stability of the molecules.  We then present the electronic and magnetic anisotropic properties of these SMMs in subsections \ref{electronic} and subsection \ref{anisotropic}, respectively. Additionally, the magnetization tunneling for these molecules is discussed in subsection \ref{spinrel}, which is important to consider when assessing how they can be applied, as discussed in subsection \ref{application}. Finally, we present the summary of our work in section \ref{summary}.

\section{Computational methods}
\label{computation}

Electronic structure calculations were performed using the all-electron NRLMOL code\cite{Pederson1990} that uses a highly optimized Gaussian basis set\cite{Porezag1999} to solve the Kohn-Sham equations using Perdew-Burke-Ernzerhof (PBE) generalized gradient approximation (GGA)\cite{Perdew1996}. First, the molecular geometries of all metallocenes were fully relaxed until the forces were below $10^{-3}$ Hartree/Bohr and the total-energy change below $10^{-6}$ Hartree. To investigate the structural stability of 4d metallocenes, we have carried out vibrational calculations. After obtaining relaxed and stable structures, SOC calculations were performed using the exact full-space approach as implemented in NRLMOL\cite{Pederson1999}, with a total energy tolerance of $10^{-6}$ Hartree. Previous studies have indicated that this approach can lead to quantitatively accurate results in molecular magnets composed of magnetic clusters linked by organic linkers. To estimate MAE and the easy axis, the SOC calculations for each metallocene are performed for 17 different quantization directions, with the polar angle $\theta$ taking five different equidistant values between $0$ and $\pi/2$ relative to the molecular axis ($\hat{z}$), and five the azimuthal angles between $0$ and $2\pi$.

\section{Results and discussion}

\subsection{Structure and structural stability}
\label{stability}

\begin{figure}[h]
    \centering
    \subfloat[\centering Full-ligand (side view)]{{\includegraphics[width=3.7cm]{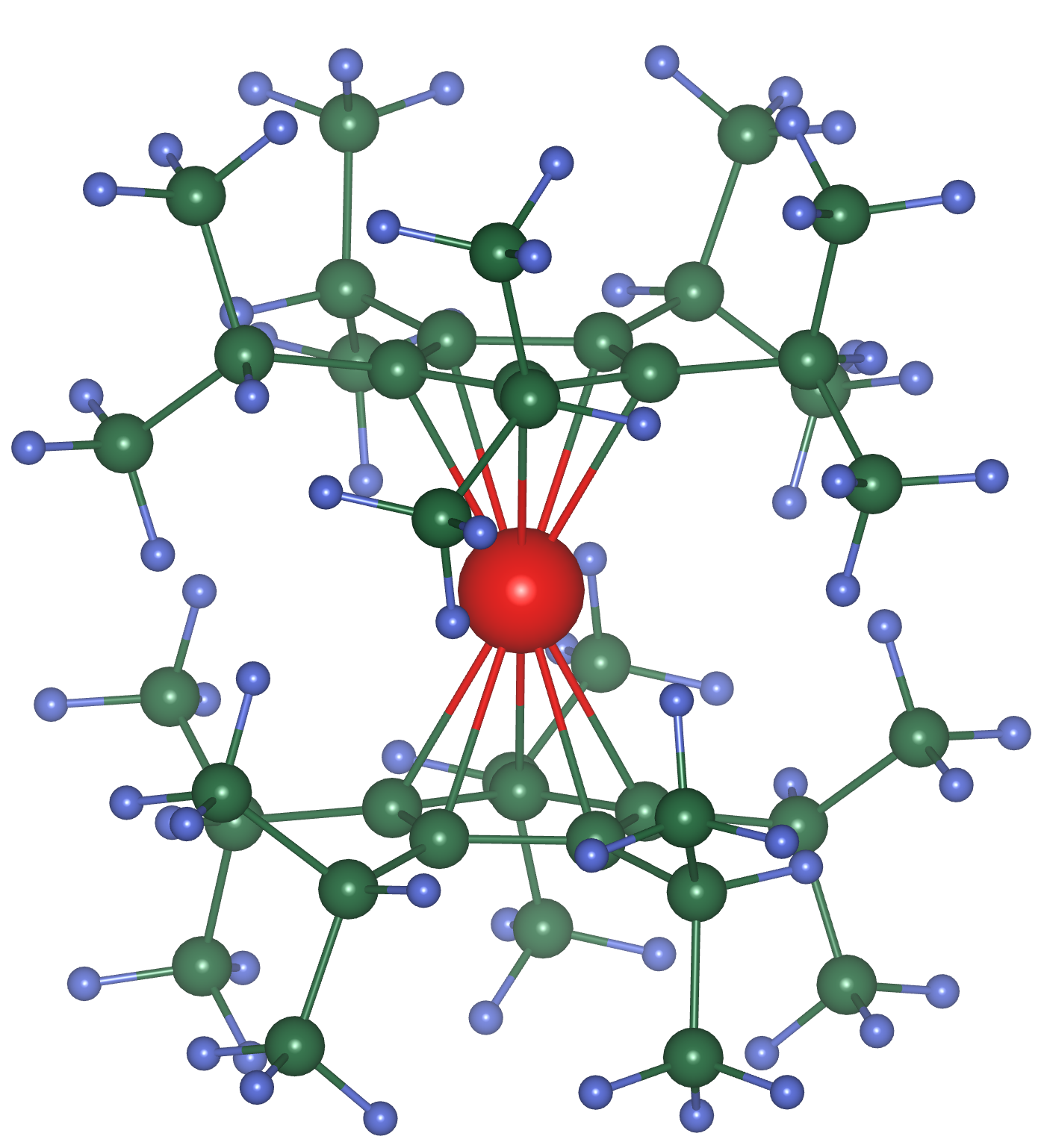}}}%
    \qquad
    \vspace{-0.3cm}
    \hspace{-0.5cm}
    \subfloat[\centering Full-ligand (top view)]{{\includegraphics[width=3.85cm]{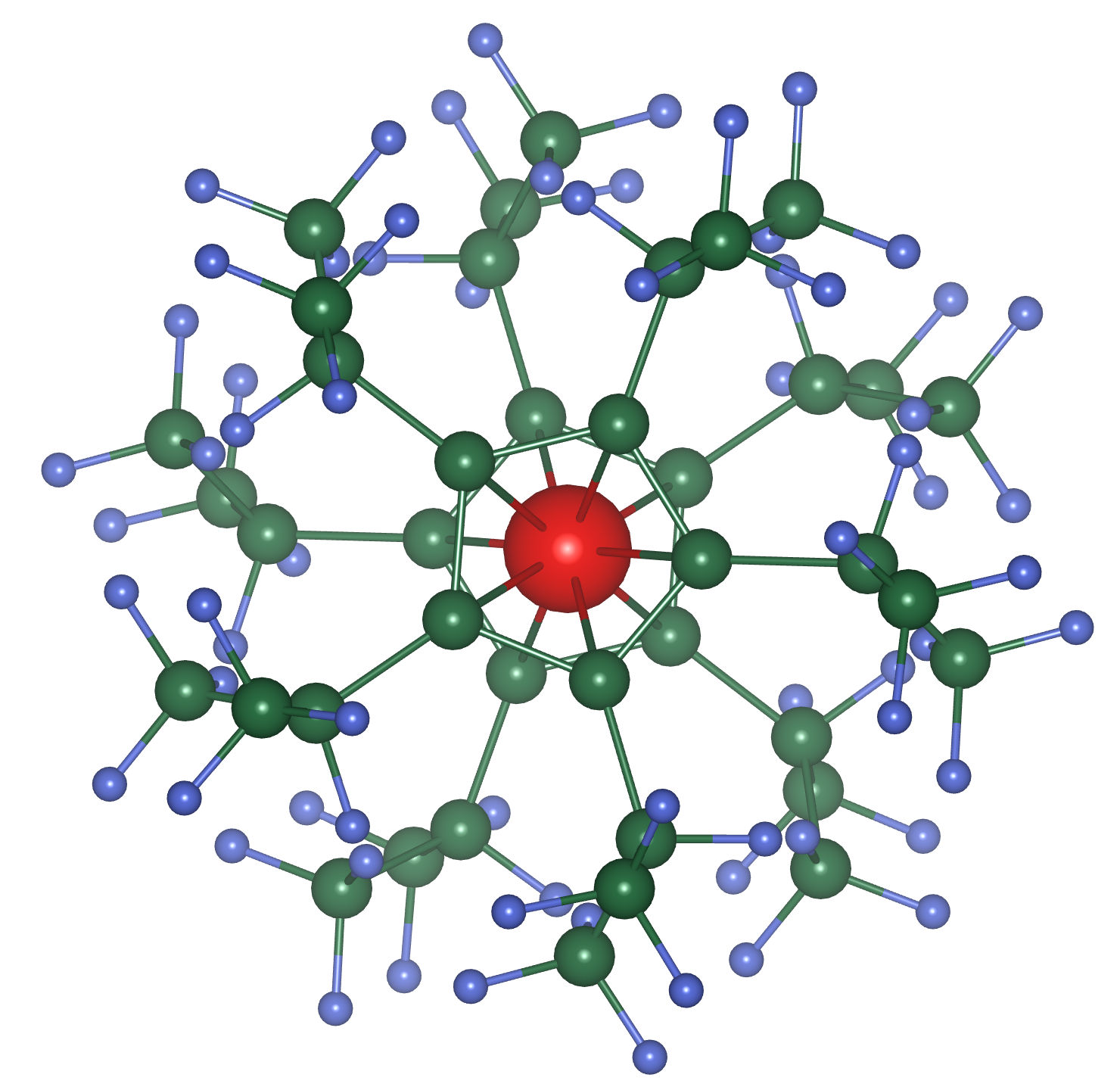}}}%
    \qquad
    \subfloat[\centering CH$_{3}$ substituent]{{\includegraphics[width=3.4cm]{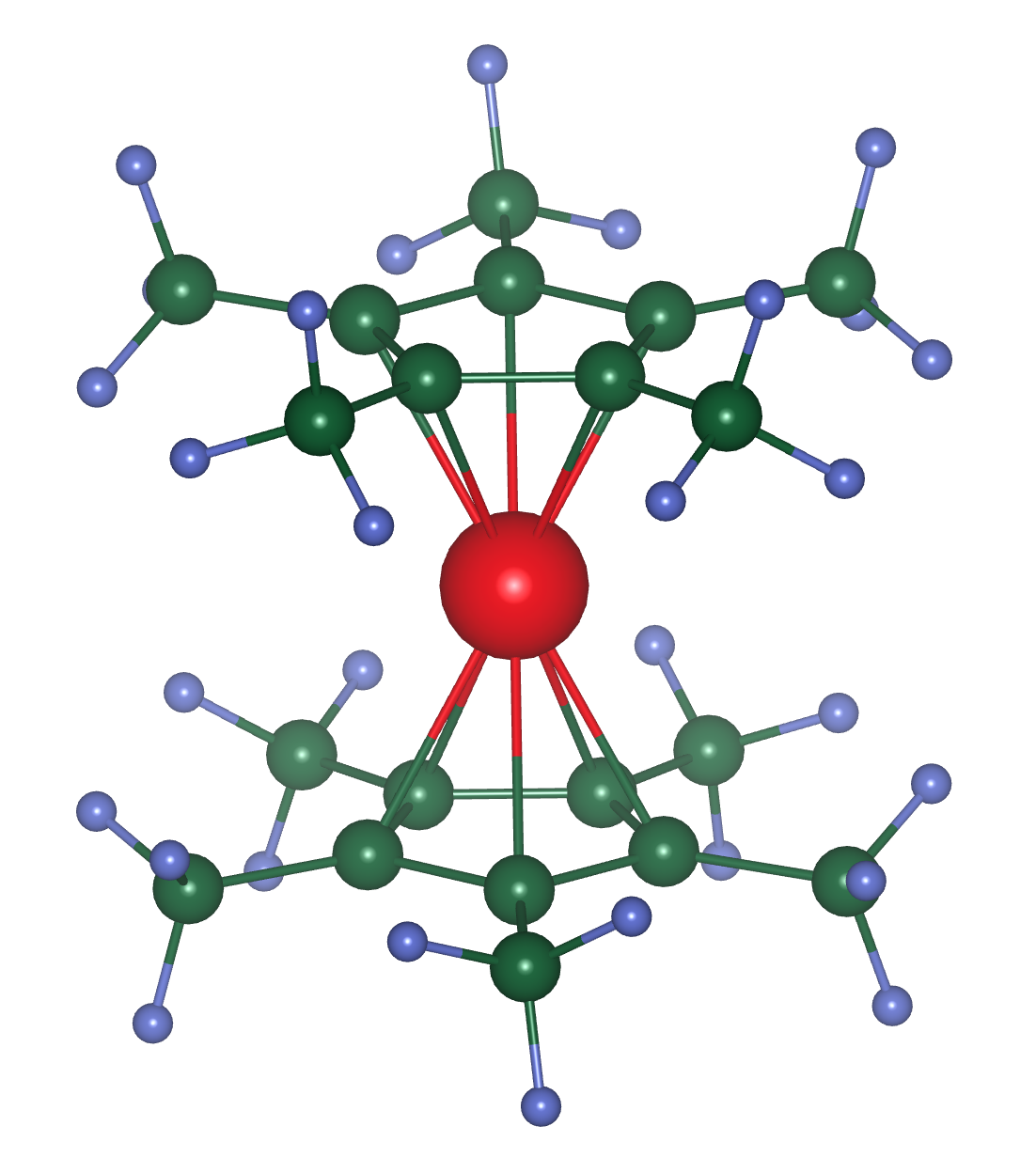}}}%
    \qquad
    \hspace{0.5cm}
    \subfloat[\centering H substituent]{{\includegraphics[width=2.8cm]{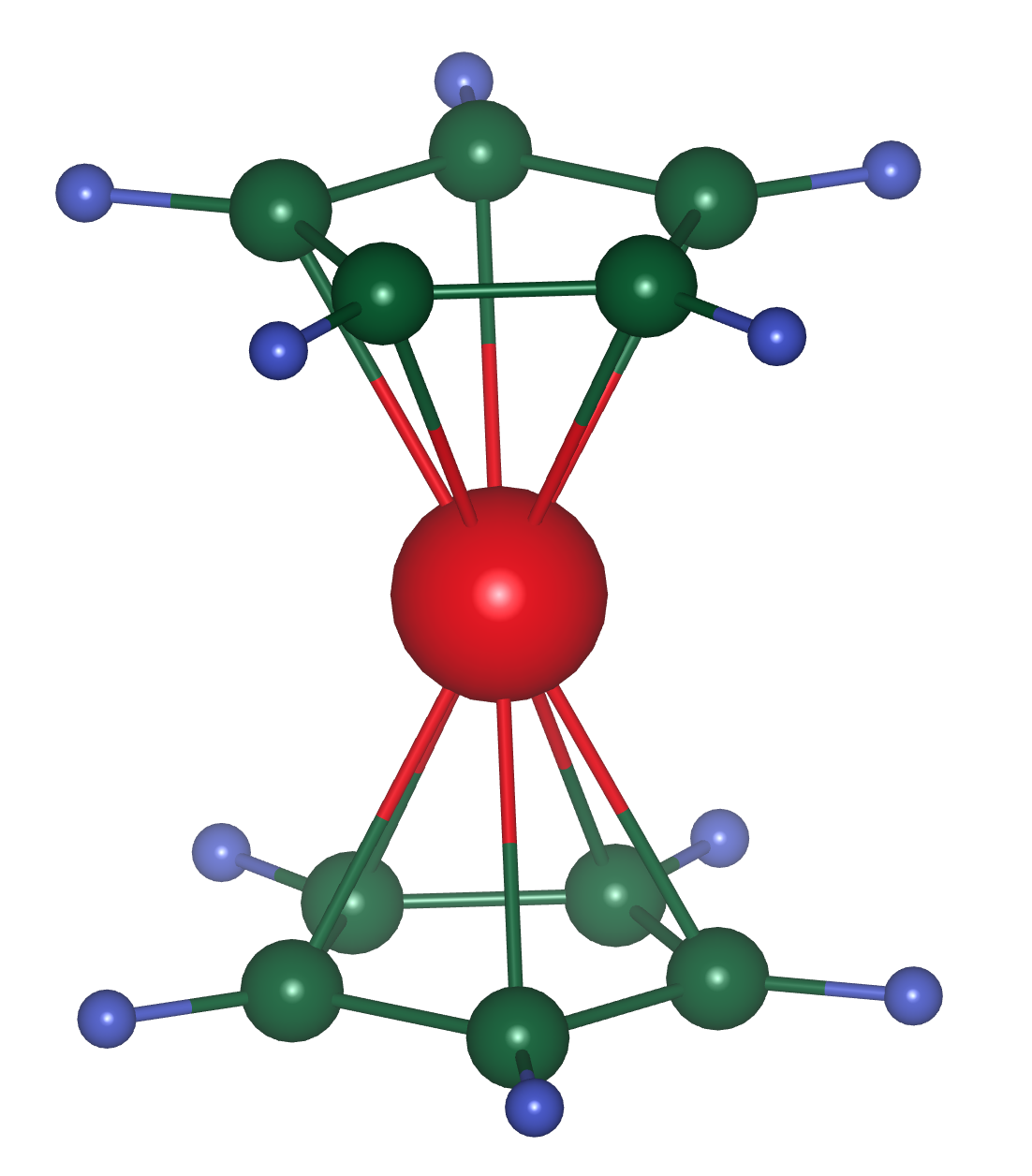}}}%
    \caption{Metallocene with transition metal elements denoted as red ball at the center. The green and blue balls correspond to carbon and hydrogen, respectively.}%
    \label{metallocene}%
\end{figure}

Metallocenes (MCp$_2$) are organometallic complexes with the general structure Cp–M–Cp, with M a transition-metal center and Cp the cyclopentadienyl anion \ce{C5H5-} (see Figs.~\ref{metallocene} a and b). These molecules have been synthesized in different ligand environments. Depending on the ligand groups these molecules can be realized in different point group symmetries such as $D_{5h}$, $D_{5d}$, or $S_{10}$. The structural data for this work was obtained from \cite{Gould2019}, which was synthesized for Dysprosium (Dy) and Terbium (Tb) as metal centers in $S_{10}$ environment. In this work, we have used the same ligand environment but with 4d and 3d transition metal elements to study their electronic and magnetic anisotropic properties. Our theoretical calculations utilized a metallocene model consisting of 111 atoms, of which only 12 atoms are inequivalent within the $S_{10}$ symmetry.

Since metallocenes containing 4d or 3d elements have not been synthesized experimentally in the same ligand environment as DyCp$_2$ and TbCp$_2$, it is essential to access their structural stability. To this end, we have calculated their vibrational modes within the harmonic approximation. The results are summarized in the TABLE \ref{tab:vib}. 
\begin{table}[ht]
\centering
\caption{Vibrational properties of 4d and 3d metallocenes with the full-ligand. The columns display calculated vibrational frequencies in cm$^{-1}$, ranging from the lowest real value $f_\text{low}$ to the highest real value $f_\text{high}$, alongside the imaginary modes (Imag. modes) obtained using DFT. $^*$M denotes the metal center of metallocenes that underwent Jahn-Teller distortion, and  M$^+$ denotes the metal center of +1 charge state metallocenes.}
\vspace{\baselineskip}
\begin{tabular}{cccc} \hline\hline
\label{tab:vib}
Metal   & $f_\text{low}$ (cm$^{-1}$) & $f_\text{high}$ (cm$^{-1}$) & Imag. modes (cm$^{-1}$) \\ \hline
4d elements&  &        &  \\ \cline {1-4}
Y      & 35.6 & 3049.0 &  --     \\
Zr     & 38.0 & 3056.2 &  --  \\
Nb     & 39.6 & 3070.7 &  --  \\
Nb$^+$ & 38.7 & 3064.9 &  119.6$i$, 120.0$i$  \\
Mo     & 34.6 & 3082.0 &  35.6$i$, 101.6$i$ \\
$^*$Mo & 31.1 & 3037.7 & --  \\
Mo$^+$ & 36.1 & 3079.0 &  --  \\
Tc     & 43.6 & 3092.6 &  --     \\
Ru     & 44.1 & 3095.4 &  --     \\
Rh     & 33.5 & 3087.9 &  1973.5$i$, 1977.1$i$\\ \hline
3d elements&  &        &  \\ \cline {1-4}
V      & 40.9 & 3087.3 &  --   \\
\multirow{2}{*}{Cr} & \multirow{2}{*}{19.5} & \multirow{2}{*}{3081.4} &53.1$i$, 148.2$i$, 185.1$i$\\
       &       &         & 1113.0$i$, 1138.3$i$\\
Cr$^+$ & 45.7 & 3089.4 &  --   \\
Mn     & 13.2 & 3073.1 &  --  \\
Fe     & 48.5 & 3104.0 &  --     \\
\multirow{2}{*}{Co} & \multirow{2}{*}{14.7} & \multirow{2}{*}{3097.6} &17.5$i$, 39.7$i$, 73.1$i$\\
       &       &         & 101.4$i$, 123.1$i$, 162.2$i$\\ \hline\hline
\end{tabular}
\end{table}

For the 4d series, YCp$_2$, ZrCp$_2$, NbCp$_2$, TcCp$_2$ and RuCp$_2$ have all real vibrational modes, confirming their stability in this ligand environment. In contrast, MoCp$_2$ and RhCp$_2$ each display two imaginary vibrational modes, indicating structural instability. This instability originates from a doubly degenerate HOMO, imposed by the molecular symmetry, but occupied by a single electron. To resolve this instability, both molecules were reoptimized without symmetry constraints. In each case, the structure undergoes a Jahn--Teller distortion that lowers the symmetry to C$_i$. These distorted molecules are denoted as $^*$MCp$_2$. The corresponding distortion energies are 180 meV for MoCp$_2$ and 82 meV for RhCp$_2$. The distortions are accompanied by only slight contractions of the molecular structure, amounting to $0.6\%$ and $0.8\%$, respectively (the distortion is quantified by calculating change in the average bond length between metal center and the five C atoms in the Cp ring). As a result, the HOMO degeneracy is lifted, yielding dynamically stable structures, as confirmed by the absence of imaginary vibrational frequencies. Interestingly, ionization of MoCp$_2$ (denoted as Mo$^+$Cp$_2$) removes the electronic degeneracy, restoring the higher molecular symmetry and resulting in a stable structure.

Among the 3d metallocenes considered in this work, VCp$_2$, MnCp$_2$, and FeCp$_2$ exhibit real vibrational modes, whereas CrCp$_2$ and CoCp$_2$ display instabilities analogous to those found for MoCp$_2$. In both cases, the instability is eliminated through a Jahn--Teller distortion. The calculated distortion energies are 52.85 meV for CrCp$_2$ and 113.23 meV for CoCp$_2$, while the associated structural contractions are only about $0.1\%$. As in the 4d series, removal of the electron occupying the degenerate HOMO restores the higher molecular symmetry.

Since the molecular model used in these calculations contains 110 ligand atoms, performing DFT calculations on such large systems is computationally expensive. Therefore, it is desirable to employ smaller ligand models that preserve the electronic properties of the full ligand, provided that the reduction in ligand size does not compromise structural stability. To investigate the effect of ligand size on the stability and electronic properties of the metallocenes, we constructed two simplified ligand models. In the first model, the ligand groups attached to each carbon atom of the Cp ring are replaced by CH$_3$ group, shown in Fig.~\ref{metallocene}c, while in the second model they are replaced by hydrogen atoms, shown in Fig.~\ref{metallocene}d. We performed vibrational and electronic structure calculations for ZrCp$_2$, NbCp$_2$, Mo$^+$Cp$_2$, VCp$_2$, and FeCp$_2$ using these reduced ligand models. In both cases, the molecular symmetry was preserved. Additionally, $^*$MoCp$_2$ and $^*$CoCp$_2$ systems with the H-ligand were optimized without symmetry constraints, and vibrational calculations were performed. The vibrational analysis for the H-ligand model is summarized in Table~\ref{tab:vib2}. 

\begin{table}[ht]
\centering
\caption{Vibrational properties of selected 4d and 3d metallocenes with the h-ligand.} 
\vspace{\baselineskip}
\begin{tabular}{cccc} \hline\hline
\label{tab:vib2}
Metal   & $f_\text{low}$ (cm$^{-1}$) & $f_\text{high}$ (cm$^{-1}$) & Imag. modes (cm$^{-1}$) \\ \hline
4d elements&  &        &  \\ \cline {1-4}
Zr     & 244.1 & 3170.6 &14.8$i$, 35.31$i$, 35.43$i$ \\
Nb     & 68.0 & 3176.5 & 23.8$i$  \\
$^*$Mo & 69.9 & 3178.2 & 43.3$i$  \\
Mo$^+$ & 18.6 & 3174.4 & 23.1$i$  \\ \hline
3d elements&  &        &  \\ \cline {1-4}
V      & 83.7 & 3185.2 &  42.1$i$ \\
Fe     & 164.2 & 3171.5 & 58.4$i$ \\
$^*$Co & 139.4 & 3178.2 &35.6$i$, 134.0$i$\\ \hline\hline
\end{tabular}
\end{table}
The results show that the 4d metallocenes, ZrCp$_2$, NbCp$_2$, $^*$MoCp$_2$, Mo$^+$Cp$_2$, and the 3d metallocenes FeCp$_2$, $^*$CoCp$_2$, which are dynamically stable with the full ligand, develop imaginary vibrational modes when H ligand is used. A similar pattern is also observed with the CH$_3$ ligand model. The appearance of imaginary modes with reduced ligands indicates that the corresponding structures are no longer true local minima on the potential energy surface. Previous theoretical studies demonstrated that FeCp$_2$ with the H-ligand in the D$_{5h}$ environment is stable, while the D$_{5d}$ environment corresponds to a transition state\cite{Zhen2003}. The S$_{10}$ symmetry of FeCp$_2$ studied in this work is a subgroup of D$_{5d}$, suggesting that the reduced-ligand models are close to transition states. In the case of VCp$_2$ and Mo$^+$Cp$_2$, imaginary modes appear only with the H-ligand, whereas all vibrational modes remain real for the CH$_3$-ligand model. These findings demonstrate that the full ligand plays a crucial role in stabilizing metallocene structures within the $S_{10}$ symmetry. 

\begin{figure}[!htb]
  \includegraphics[width=1\linewidth]{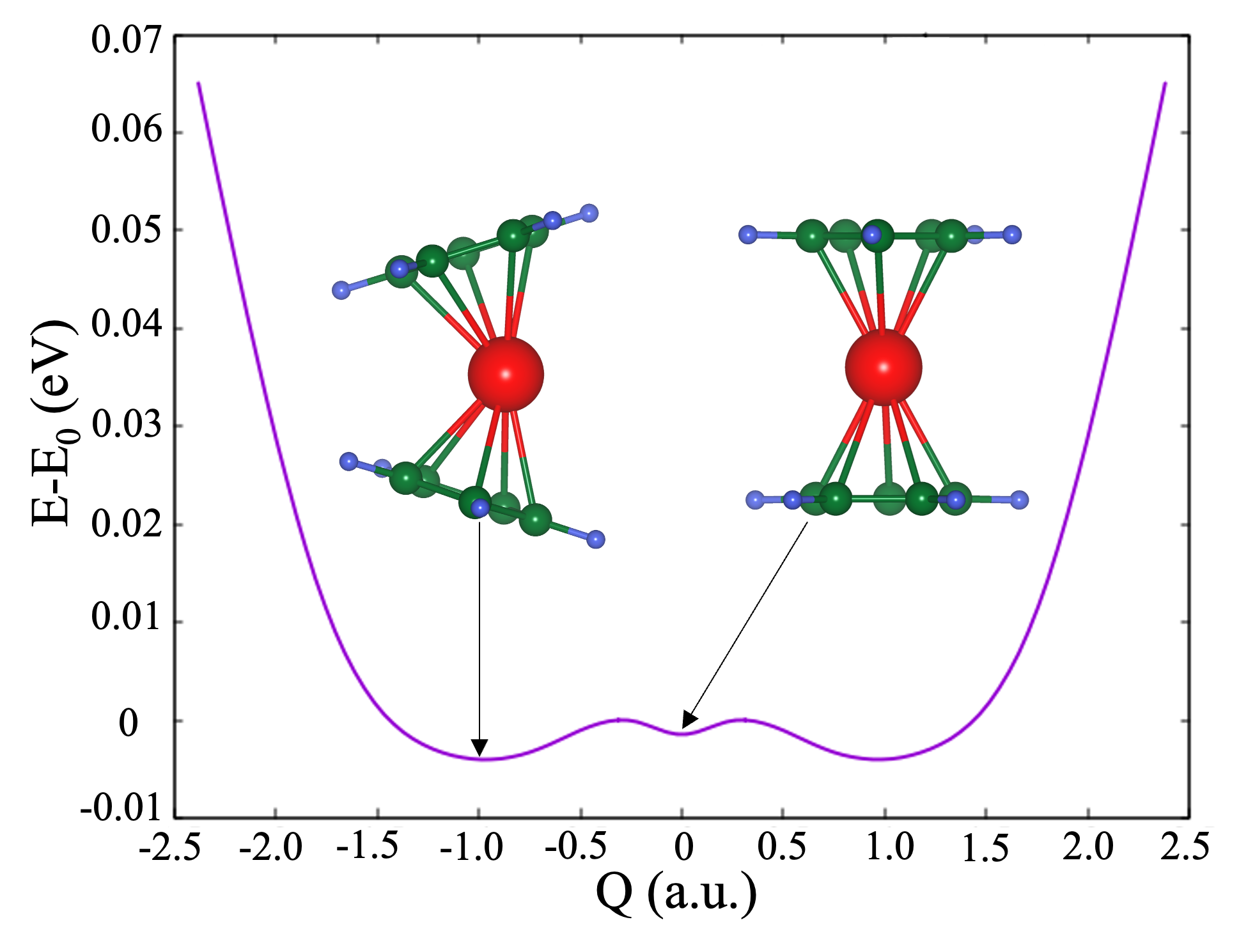}
  \caption{Double-well potential energy surface of ZrCp$_2$ for different displacements Q along one imaginary vibrational mode and the corresponding optimized ground-state structure.}
  \label{fig:Zr_bend}
\end{figure}

To investigate the origin of imaginary vibrational modes, we calculated the potential energy surface (PES) of ZrCp$_2$ with the H-ligand as a function of the displacement coordinate, (Q), along the eigenvector of the imaginary mode (ZrCp$_2$ is chosen for its larger number of imaginary modes). The resulting PES is shown in Fig.~\ref{fig:Zr_bend}. It is evident from the figure that the symmetric structure (with Q=0) corresponds to a saddle point on the PES, resulting in an imaginary vibrational mode. To reach the global minima at Q $\sim \pm$ 1, we relaxed the system without any symmetry constraint and found that the molecule undergoes a significant structural distortion. This distortion can be understood by comparing the ligand groups in Fig.~\ref{metallocene}a (corresponding to the full-ligand) and $d$ (corresponding to the H-ligand), which show that the ligand atoms between two Cp rings are close to each other in the full-ligand, preventing the structure from bending.      

In addition, we investigated how the reduced model of ligands affects the electronic properties of the systems. Our results show that, as long as the $S_{10}$ molecular symmetry is preserved (i.e., the structure remains at a local minimum), key electronic properties, including energy-level ordering and the magnetic anisotropy energy, remain essentially unchanged relative to the full-ligand structure (discussed in detail in Sec.~\ref{electronic}), despite the appearance of imaginary vibrational modes. In contrast, the structure corresponding to the global minimum undergoes substantial distortions, as shown in Fig.~\ref{fig:Zr_bend}.

These results demonstrate that reduced-size ligand models can reliably reproduce the electronic structure and magnetic anisotropy of high-symmetry configurations, supporting the validity of previous studies employing simplified ligand environments\cite{Zhen2003,Nawa2016,Mukhopadhyaya2024,Lunghi2019,Guo2017}. However, because reduced ligands significantly modifies the vibrational spectrum, preserving the full-ligand structure is essential to accurately describe properties that depend on molecular vibrations, such as electron-vibron and spin-vibron coupling.

\subsection{Electronic properties}
\label{electronic}

In this subsection, we discuss how the electronic structure of metallocenes evolves as the number of $d$ electrons increases, since the distribution of electrons among the $d$ orbitals and their interaction with the ligand field determine the electronic and magnetic properties of these molecules.


To calculate the electronic properties, we have used the structures corresponding to the full-ligand described in subsection~\ref{stability}. All calculations are performed by preserving the $S_{10}$ molecular symmetry, except for those that underwent Jahn-Teller distortion. Across the 4d series, the M–C\(_\text{ring}\) distance contracts from YCp$_2$ to RuCp$_2$ (2.73~\AA\ to 2.27~\AA; Table~\ref{tab:Elec_data}), consistent with the increase in effective nuclear charge across the series, which strengthens the metal-ligand interaction. A similar but less systematic trend is observed for the 3d-series, where there is an increase in the  M–C\(_\text{ring}\) distance from VCp$_2$ to MnCp$_2$ and then a decrease for FeCp$_2$ and $^*$CoCp$_2$.

\begin{table}[ht]
\centering
\caption{Electronic properties of 4d and 3d metallocenes as a function of the $d$-electron occupation ($d$ occ.) of isolated atoms. The remaining columns correspond to calculated magnetic moment (Mom.), metal-ring carbon distance (M-C$_\mathrm{ring}$), and HOMO-LUMO gap (HL gap) obtained using DFT. $^*$M denotes the metal center of metallocenes that underwent Jahn-Teller distortion, and  M$^+$ denotes the metal center of +1 charge state metallocenes.}
\vspace{\baselineskip}
\begin{tabular}{ccccc} \hline\hline
\label{tab:Elec_data}
Metal   & $d$ occ. & Mom. ($\mu_B$) & M-C$_\mathrm{ring}$ (Å) & HL gap (eV) \\ \hline
4d elements&  &              &                         &\\ \cline {1-5}
Y      & 1  &    1           & 2.73                    & 0.19            \\
Zr     & 2  &    2           & 2.56                    & 0.07            \\
Nb     & 4  &    3           & 2.48                    & 1.57            \\
Nb$^+$ & 3  &    2           & 2.48                    & 0.29            \\
$^*$Mo & 5  &    2           & 2.39                    & 0.51            \\
Mo$^+$ & 4  &    3           & 2.41                    & 1.64            \\
Tc     & 5  &    1           & 2.29                    & 0.15            \\
Ru     & 7  &    0           & 2.27                    & 3.27            \\
$^*$Rh & 8  &    1           & 2.37                    & 0.34            \\ \hline
3d elements&  &              &                         &\\ \cline {1-5}
V      & 3  &    3           & 2.37                    & 1.95            \\
$^*$Cr & 5  &    4           & 2.43                    & 0.25            \\
Cr$^+$ & 4  &    3           & 2.32                    & 2.43            \\
Mn     & 5  &    5           & 2.48                    & 0.91            \\
Fe     & 6  &    0           & 2.15                    & 2.20            \\
$^*$Co & 7  &    1           & 2.23                    & 0.05            \\ \hline\hline
\end{tabular}
\end{table}

\begin{figure*}[t]
  \includegraphics[width=0.95\linewidth]{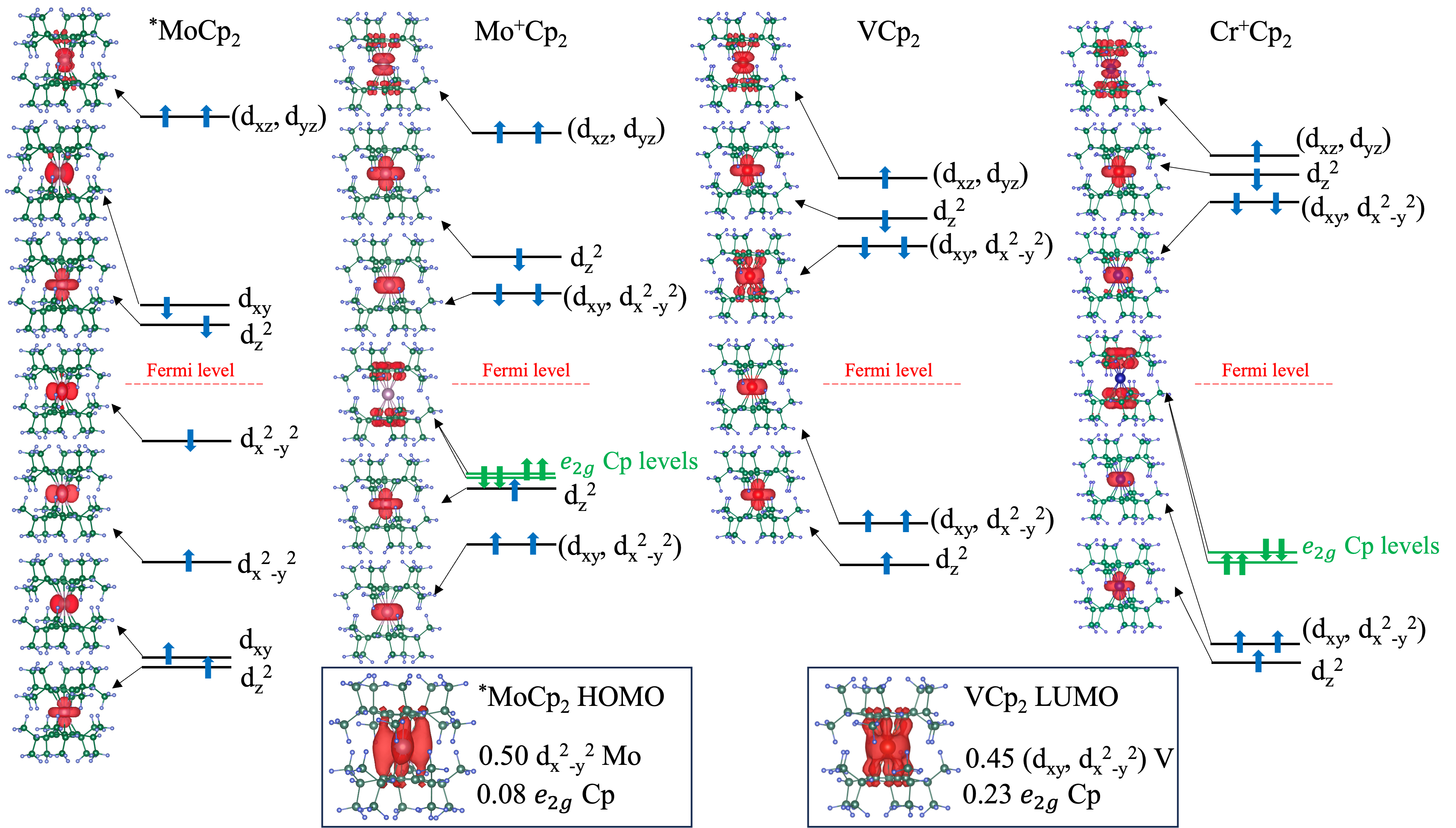}
  \caption{Schematics of energy levels of $^*$MoCp$_2$, Mo$^+$Cp$_2$, VCp$_2$ and Cr$^+$Cp$_2$ metallocenes around the Fermi level and the corresponding spatial distribution of molecular orbitals (the same isosurface cutoff is used for all plots). In the inset the HOMO of $^*$MoCp$_2$, and the LUMO of VCp$_2$ are replotted with smaller cutoff to show the fractional contribution of the Cp$_2$ $\pi$ orbitals.}
  \label{fig:Levels}
\end{figure*}

To understand the electronic structure near the Fermi level,  we first discuss the crystal-field splitting of the metal $d$ orbitals. Figure~\ref{fig:Levels} shows the real-space distributions of representative molecular orbitals for selected metallocenes. 
For the metallocenes that preserve the $S_{10}$ symmetry, the crystal field splits the five $d$ orbitals into three groups: a nondegenerate $d_{z^2}$ orbital and two doubly degenerate pairs, $(d_{xz},d_{yz})$ and $(d_{x^2-y^2},d_{xy})$. The carbon atoms of the cyclopentadienyl (Cp) ligands form $\pi$ molecular orbitals that hybridize with the metal $d$ orbitals. In particular, the $e_{1g}$ ligand orbitals interact with the $(d_{xz},d_{yz})$ orbitals, increasing their energy. In contrast, the $e_{2g}$ orbitals hybridize with the $(d_{x^2-y^2},d_{xy})$ orbitals, lowering their energy. Since the $d_{z^2}$ orbital has little direct overlap with the ligand orbitals, it is expected to remain the lowest in energy.

To complement the real-space visualization of the molecular orbitals, we performed a quantitative orbital projection analysis. For most of the investigated metallocenes, the HOMO and LUMO are predominantly derived from the transition-metal $d$ orbitals, which contribute between 40\% and 70\% of the total orbital character. The $\pi$ orbitals of the Cp rings contribute less than 40\% in most cases, indicating that the molecular orbitals near the Fermi level retain a predominantly metal-$d$ character despite the metal-ligand hybridization. There are, however, some exceptions. In MnCp$_2$, the HOMO has mainly Cp $\pi$ character, with about 60\% contribution from the ligands and only 29\% from the Mn $d$ orbitals. Similarly, the HOMOs of Mo$^+$Cp$_2$ and Cr$^+$Cp$_2$ are largely composed of Cp $\pi$ orbitals, but the contribution from the metal center is negligible.

Figure~\ref{fig:Levels} also illustrates the effect of Jahn-Teller distortion on the molecular orbitals. In $^*$MoCp$_2$, the distortion lifts the degeneracy of the $(d_{x^2-y^2},d_{xy})$ doublet, removing the single occupation of degenerate orbitals and lowering the total energy. After ionization, the molecular symmetry is restored in Mo$^+$Cp$_2$, and the degeneracy of the doublet is recovered. In both systems, the $(d_{xz},d_{yz})$ orbitals strongly hybridize with the $e_{1g}$ orbitals of the Cp rings, shifting these antibonding states to higher energies.

The comparison between $^*$MoCp$_2$ and Mo$^+$Cp$_2$ shows that ionization also modifies the HOMO-LUMO gap and the ordering of the frontier molecular orbitals. VCp$_2$ and Cr$^+$Cp$_2$ display similar $d$-electronic structures; however, the HOMO-LUMO gap is larger for Cr$^+$Cp$_2$. These differences in the electronic structure will be discussed further in connection with the magnetic anisotropy in the next subsection.

\begin{figure*}[htb]
  \includegraphics[width=1\linewidth]{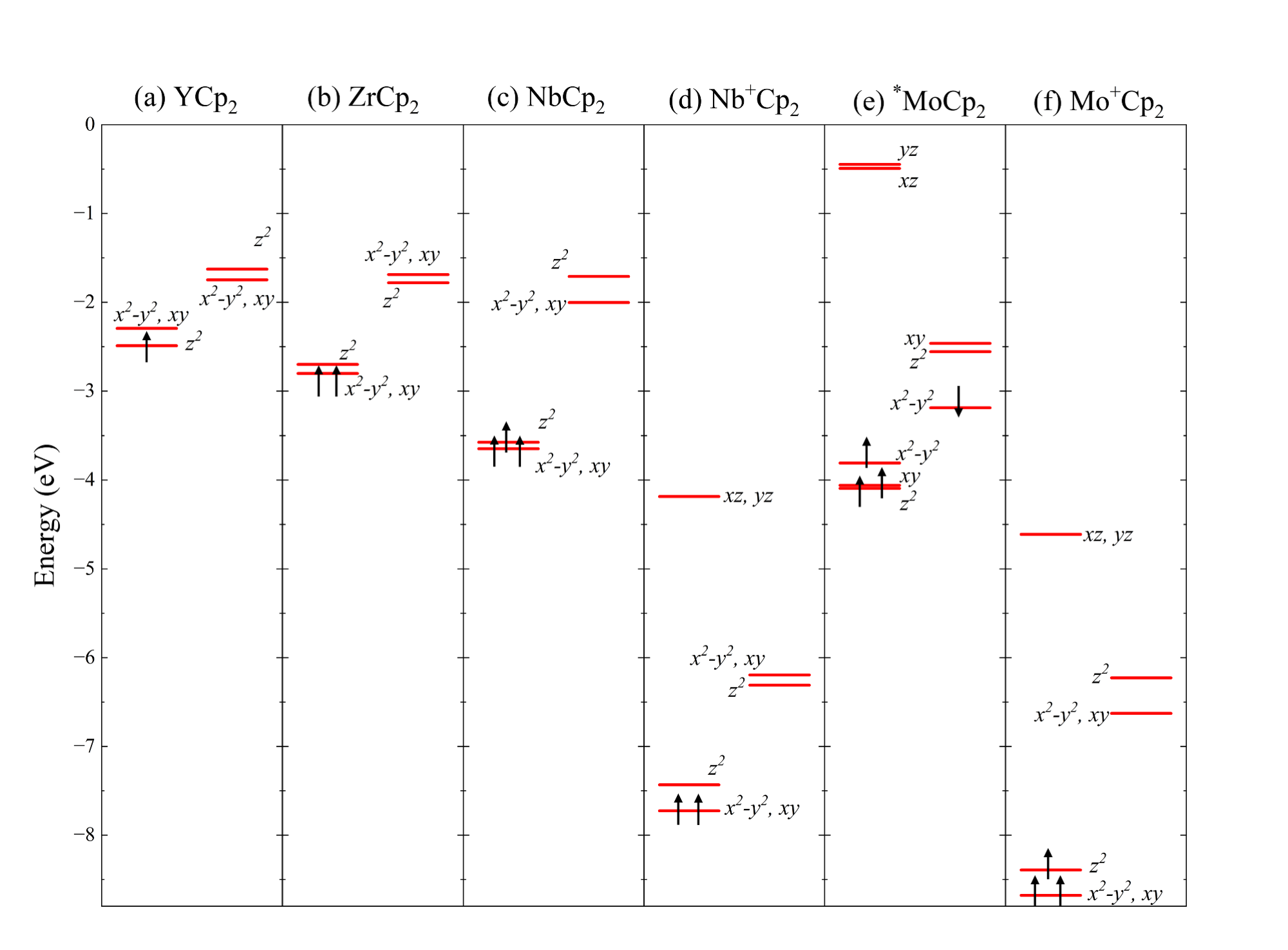}
  \caption{Energy level diagrams of 4d metallocenes showing the splitting and occupation of the metal $d$ orbitals. The left and right columns correspond to the spin-up and spin-down channels, respectively. Arrows indicate the occupation of each orbital. The zero energy corresponds to the vacuum energy.}
  \label{fig:MO4d}
\end{figure*}

\begin{figure*}[htb]
  \includegraphics[width=1\linewidth]{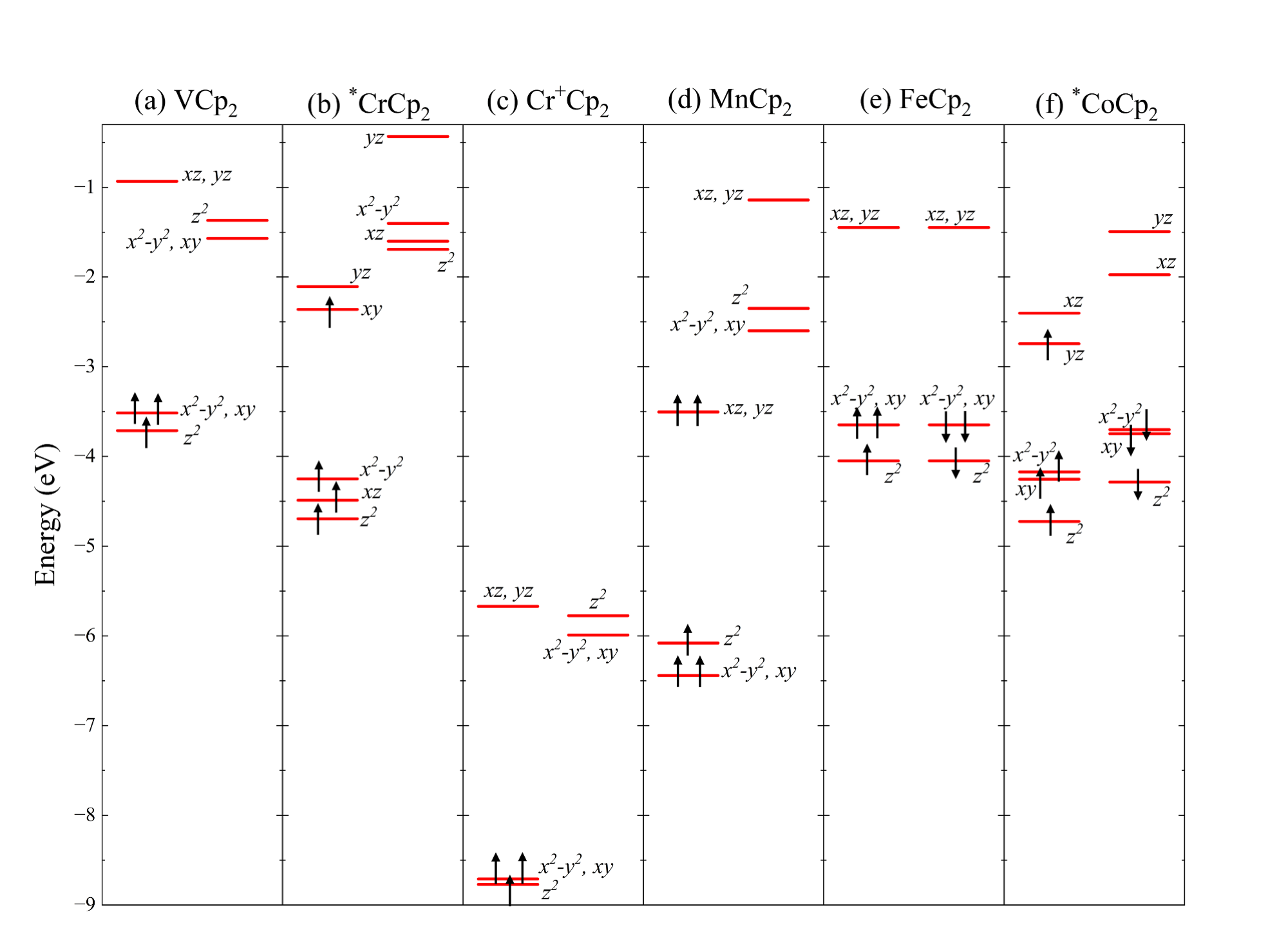}
  \caption{Energy level diagrams of 3d metallocenes showing the splitting and occupation of the metal $d$ orbitals. The left and right columns correspond to the spin-up and spin-down channels, respectively. Arrows indicate the occupation of each orbital. The zero energy corresponds to the vacuum energy.}
  \label{fig:MO3d}
\end{figure*}

\begin{figure*}[!htb]
  \includegraphics[width=0.7\linewidth]{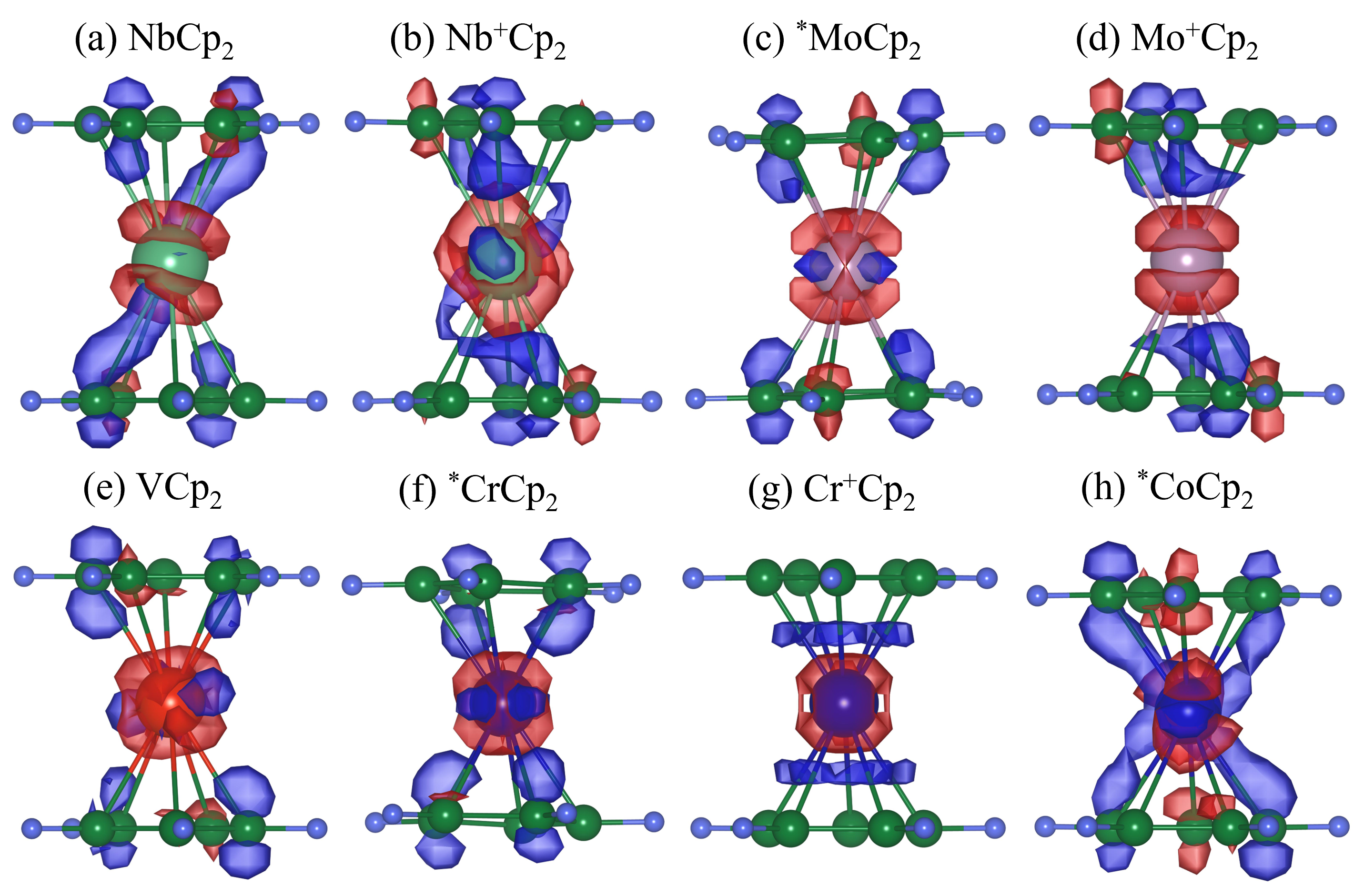}
  \caption{Charge-density difference $\Delta\rho=\rho_{MCp_{2}}-\rho_{M}-\rho_{Cp_{2}}$ for (a) NbCp$_2$, (b) NbCp$_2$, (c) $^*$MoCp$_2$, (d) Mo$^+$Cp$_2$, (e) VCp$_2$, (f) $^*$CrCp$_2$, (g) Cr$^+$Cp$_2$ and (h) $^*$CoCp$_2$. The blue (positive) and red (negative) regions indicates charge accumulation and depletion, respectively.}
  \label{fig:ch_den_diff}
\end{figure*}

The ordering of molecular $d$ energy-levels for the 4d and 3d metallocene series is presented on a common energy scale in Figs.~\ref{fig:MO4d} and \ref{fig:MO3d}, respectively. Although TcCp$_2$, RuCp$_2$, and $^{*}$RhCp$_2$ were also calculated, they are omitted from Fig.~\ref{fig:MO4d} for clarity. In general, formation of the cationic state leads to a downward shift of the single-particle energy levels. This shift arises from the reduction of electron–electron Coulomb repulsion upon removal of an electron from the metal center, which enhances the effective nuclear attraction experienced by the remaining electrons. Moreover, the greater spatial extent of the 4d orbitals compared to the 3d orbitals results in stronger metal–ligand hybridization in the 4d metallocene series. Consequently, the 4d metallocenes exhibit a larger crystal-field splitting, as illustrated in Figs.~\ref{fig:MO4d} and \ref{fig:MO3d}. This is further supported by the calculated charge-density differences for selected MCp$_2$ systems (Fig.~\ref{fig:ch_den_diff}), defined as $\Delta\rho=\rho_{MCp_{2}}-\rho_{M}-\rho_{Cp_{2}}$. This charge redistribution indicates a transfer of electronic charge from the metal center toward the ligand environment. The effect is more pronounced for the 4d metallocenes, consistent with their stronger metal-ligand hybridization.

The $d$-energy level structure shown in Figs.~\ref{fig:MO4d} and \ref{fig:MO3d} can be used to understand the calculated magnetic moments of the metallocenes tabulated in Table~\ref{tab:Elec_data}. For example, the magnetic moment of the Nb-metallocene can be understood from the ordering of the corresponding energy level depicted in Fig.~\ref{fig:MO4d}c. Since three of the $d$ electrons occupy the up-spin states and only one is in the down-spin state, the total moment is 2$\mu_B$ as reported in Table~\ref{tab:Elec_data}. It is also evident from Fig.~\ref{fig:MO4d}d that when an electron is removed from the highest occupied down-spin level, the moment increases to 3$\mu_B$. The calculated magnetic moments of other metallocenes can also be understood from the electronic structure diagrams. Comparing Table~\ref{tab:Elec_data} and Figs.\ref{fig:MO4d} and \ref{fig:MO3d}, we note that the electron count of the $d$ level generally differs from that of the isolated atom. This discrepancy arises from the delocalization of some $d$ electrons within the ligand environment.   

Overall, the electronic structure shows that changes in molecular symmetry and charge state modify the orbital ordering and the metal-ligand hybridization. As shown in the next subsection, these changes directly affect the magnetic anisotropy of the metallocenes.

\subsection{Magnetic anisotropic properties}
\label{anisotropic}

One of the objectives of this work is to understand how the electronic structure of metallocenes determines their magnetic anisotropy and the stability of their magnetic states. The potential of these molecules for information storage depends critically on the uniaxial magnetic anisotropy barrier that separates their two degenerate ground states. These states represent the fundamental quantum configurations that can encode binary information, 0 and 1. A sufficiently large MAE barrier, denoted by $\Delta E_\text{MAE}$, stabilizes these states against thermal fluctuations and suppresses spontaneous magnetization reversal. Consequently, the maximum operating temperature of the memory device increases with the height of this barrier. It is calculated from the difference in the SOC energy for the easy and hard magnetization axes. In this subsection, we analyze two key requirements that govern the suitability of metallocenes as memory elements, namely the degree of magnetic uniaxiality and the magnitude of the magnetic anisotropy barrier.

We first discuss the DFT results of our magnetic anisotropy calculations. Table~\ref{tab:MAE} summarizes the MAEs and the corresponding easy-axis directions for both 4d and 3d metallocenes. The table also reports the axial (D) and transverse (E) zero-field splitting parameters, which characterize the anisotropic spin interactions arising in systems with more than one unpaired electron. A more detailed discussion of these parameters along with the effective spin model is provided in subsection \ref{spinrel}. The calculated MAEs range from a negligible 0.05 K for YCp$_2$ to a maximum of 60.4 K for Mo$^+$Cp$_2$. The calculations further show that most metallocenes exhibit an easy-plane magnetic anisotropy perpendicular to the molecular axis. 

\begin{table}[ht]
\centering
\caption{Magnetic properties of 4d and 3d metallocenes from DFT+SOC calculations. The columns correspond to calculated magnetic anisotropy energy ($\Delta E_\text{MAE}$), easy magnetization orientation M$_\text{easy}$, and $D$(K) and $E$(K) are the axial and transverse zero-field anisotropy parameters, respectively. “001” denotes uniaxial magnetic anisotropy (along the molecular axis) and “easy-plane” indicates the anisotropy transverse to the molecular axis. All energies are given in units of Kelvin (K).}
\vspace{\baselineskip}
\begin{tabular}{cccccc} \hline\hline
\label{tab:MAE}
Metal&Mom. ($\mu_B$)&$\Delta E_\text{MAE}$ (K)&M$_\text{easy}$ & $D$(K)&$E$(K)\\ \hline
\multicolumn{2}{l}{4d elements}&             &      &         & \\ \cline {1-6}
Y      & 1  &    0.05     & 001        &  --     & --   \\
Zr     & 2  &    21.7     & easy-plane &  22.01  & 0.00 \\
Nb     & 3  &    32.6     & easy-plane &  14.71  & 0.00 \\
Nb$^+$ & 2  &    46.7     & easy-plane &  47.24  & 0.00 \\
$^*$Mo & 2  &    19.9     & 001        &  -20.80 & -0.01\\
Mo$^+$ & 3  &    60.4     & easy-plane &  27.11  & 0.00 \\
Tc     & 1  &    0.5      & 001        &  --     & --   \\
Ru     & 0  &    --       & --         &  --     & --   \\
$^*$Rh & 1  &    22.5     & 001        &  --     & --   \\ \hline
\multicolumn{2}{l}{3d elements}&             &      &         &\\ \cline {1-6}
V      & 3  &    3.7      & easy-plane &  1.65   & 0.00 \\
$^*$Cr & 4  &    7.8      & 001        &  -1.26  & -0.76\\
Cr$^+$ & 3  &    6.32     & easy-plane &  2.83   & 0.00 \\
Mn     & 5  &    2.2      & 001        &  -0.42  & 0.00 \\
Fe     & 0  &    --       & --         &  --     &  --  \\
$^*$Co & 1  &    6.4     & 001         &  --     &  --  \\ \hline\hline
\end{tabular}
\end{table}
To elucidate the physical origin and the easy axis of the magnetic anisotropy, we carried out a qualitative analysis of the electronic structure across the 4d and 3d metallocenes, we utilize the second-order perturbation theory to estimate the contribution of SOC to the ground-state energy for different quantization axes given by
\begin{equation}
    \Delta E_{\text{SO}} = -\sum_j \frac{|\bra{j} \hat{H}_{\text{SO}} \ket{\text{gr}}|^2}{\varepsilon_j - \varepsilon_{\text{gr}}}
\label{E_so}
\end{equation}
where the SOC Hamiltonian $H_\text{SO}$ is expressed in terms of ladder operators $L_+$ and $L_-$ as,

\begin{align}
    \hat{H}_{\text{SO}} &= \lambda (L_x S_x + L_y S_y + L_z S_z) \nonumber \\
                        &= \lambda (L_z S_z + \frac{1}{2}L_+S_- + \frac{1}{2}L_-S_+)
\label{H_so}
\end{align}
Here $\lambda$ is the effective SOC parameter that determines the strength of the interaction between the electronic spin $\vec{S}$ and orbital angular momenta $\vec{L}$. To understand the angular dependence of the SOC energy and the corresponding easy axis, we align the magnetization along an arbitrary direction {\bf n}($\theta$), where $\theta$ is the polar angle relative to the molecular axis ($C_5$ axis). Due to the azimuthal symmetry of the molecule, we neglected the dependence on the azimuthal angle $\phi$ and focused only on the polar angle $\theta$. The SOC Hamiltonian can then be expressed as
\begin{align}
    \hat{H}_{\text{SO}} &= \lambda S_z' \left( L_z \cos\theta + \frac{1}{2} L_+ \sin\theta + \frac{1}{2} L_- \sin\theta \right) \nonumber \\
    &+ \frac{\lambda}{2} S_+' \left( -L_z \sin\theta - L_+ \frac{\sin^2\theta}{2} + L_- \frac{\cos^2\theta}{2} \right) \nonumber \\
    &+ \frac{\lambda}{2} S_-' \left( -L_z \sin\theta + L_+ \frac{\cos^2\theta}{2} - L_- \frac{\sin^2\theta}{2} \right)
\label{H_so2}
\end{align}

We now analyze the angular dependence of the SOC energy using Eq.~\ref{E_so}, taking the $d$ energy levels shown in Figs.~\ref{fig:MO4d} and \ref{fig:MO3d} as the basis states. Since the largest contribution to the SOC energy comes from the states near the Fermi energy, we can estimate the angular dependence by considering only the highest occupied and lowest unoccupied $d$ states that can be coupled by $H_\text{SO}$. The advantage of expressing $H_\text{SO}$ with explicit dependence on the angle is obvious: i) If the two states are of the same spin species and of the same orbital character i.e the change in the angular momentum projection $\Delta m_l=0$, then only non-zero contribution comes from the $S_z'L_zcos\theta$ term in the Hamiltonian Eq.~\ref{H_so2}, ii) if $\Delta m_l=0$ and the two states belong to opposite spin species ($\Delta m=\pm 1$), then the $S_{\pm}'L_zsin\theta$ term will give the non-zero contribution, iii) if $\Delta m = 0$ but $\Delta m_l=\pm 1$, then $S_z'L_{\pm}cos\theta$ term is non-zero, and iv) if both $\Delta m_l,\Delta m =\pm 1$, the the non-zero contribution comes from one of the remaining terms of the Hamiltonian. 

In order to illustrate the procedure described above, we first consider ZrCp$_2$. For this metallocene, both HOMO and LUMO have the same orbital character but opposite spins, which satisfy condition (ii) above. Therefore, the angular dependence of the SOC energy is $\Delta E_\text{SO}(\theta) \sim -|sin\theta|^2$, which, in turn, implies an easy-plane magnetic anisotropy as obtained in the DFT calculations. Condition (ii) is also satisfied by both NbCp$_2$ and Nb$^+$Cp$_2$, resulting in an easy-plane magnetic anisotropy. In the case of neutral $^*$MoCp$_2$ with broken $S_{10}$ symmetry due to Jahn-Teller distortion, the ground state is $d_{x^2-y^2}\downarrow$. The LUMO level $d_{z^2}\downarrow$ does not mix with the HOMO by $H_\text{SO}$, since they differ by $\Delta m_l = 2$. However, the next unoccupied level $d_{xy}\downarrow$ can couple with the HOMO by $H_\text{SO}$. Since these two states satisfy condition (i), $\Delta E_\text{SO}(\theta) \sim -|cos\theta|^2$. Consequently, the easy-axis is along the molecular axis, which corresponds to uniaxial magnetic anisotropy. On the other hand, the cation Mo$^+$Cp$_2$, in which $S_{10}$ symmetry is restored, satisfies condition (ii), resulting in easy-plane magnetic anisotropy. Following the same procedure, one can explain if a magnetic system has easy-axis or easy-plane magnetic anisotropy. Note that the procedure described here is general and valid for any magnetic system, finite or periodic\cite{Shahid2022}. 

For Jahn-Teller distorted systems, the axial symmetry is broken and consequently the SOC energy exhibits azimuthal dependence. To study this dependence with more precision, we have performed the SOC calculations with 11 different $\theta$ angles  between 0 and $\pi/2$ and for each $\theta$, we have calculated the SOC energies for 20 different $\phi$ angles between 0 and $2\pi$. Because of a very small non-zero transverse anisotropy, this azimuthal dependence is negligible in $^*$MoCp$_2$. But for $^*$CrCp$_2$, in which transverse anisotropy is large, our calculations show that it develops an anisotropy barrier of 6 K within the xy-plane, with the lowest in-plane energy for the quantization axis along the $\pm$ y-axis ($\theta=\pi/2$ and $\phi=\pi/2$) and the largest in-plane energy is along the $\pm$ x-axis. However, in both distorted structures, the easy magnetization axis is along the molecular axis (z-direction).   

From Table~\ref{tab:MAE}, we note that there is no obvious pattern in MAE for different metallocenes, suggesting a complex dependence of MAE on their system parameters. To address this issue, we focus on Eq.~\ref{E_so}, which shows that the SOC energy depends on the SOC parameter $\lambda$, the energy gap between the occupied and unoccupied $d$ levels, and the strength of the matrix element of $H_\text{SO}$ between these states. The MAE results from a complex interplay among these three factors. To further elucidate, we consider Mo$^*$Cp$_2$ and Mo$^+$Cp$_2$. The SOC parameter for neutral Mo$^*$Cp$_2$ is about $\lambda \sim $ 86 meV, but for Mo$^+$Cp$_2$, $\lambda \sim $ 102 meV\cite{Koseki2019}. Although $\lambda$ is larger for cation, the separation between the highest occupied and lowest unoccupied $d$-level is also larger, contributing to the lowering of SOC energy. To calculate MAE, all matrix elements of $H_\text{SO}$ must be considered, which requires detailed knowledge of the spatial distribution of orbitals that are coupled by the Hamiltonian.

\subsection{Magnetization tunneling in metallocenes}
\label{spinrel}

The magnetic anisotropy discussed above establishes the energy barrier separating spin states with opposite spin projections. However, the stability of these states is not determined solely by the barrier height, which governs the thermally activated relaxation processes. Additional spin-relaxation mechanisms, such as quantum tunneling of magnetization and coupling to environmental degrees of freedom, can significantly reduce the effective stability of the magnetic states, even in systems possessing large magnetic anisotropy barriers. To assess the robustness of the magnetic states and quantify the relevant relaxation pathways, we map the magnetic properties onto an effective spin Hamiltonian\cite{atanasov2015}
\begin{align}
    \hat{H}_{\text{Spin}} &= D(S_{z}^2 - \frac{1}{3}S(S+1)) + E (S_{x}^2-S_{y}^2) + g\mu_{B}\vec{H}\cdot\vec{S}
\label{H_spin}
\end{align}
where $\vec{S}$ is the total spin of the molecule, and $D$ and $E$ are the axial and transverse zero-field splitting parameters, respectively, which characterize the interaction of the spin with surrounding orbital environments and depend on the crystal field symmetry. The last term represents the Zeeman Hamiltonian, where $\vec{H}$ is an external magnetic field, $\mu_{B}$ the Bohr's magneton, and $g=2$ the Land\'{e} $g$-factor of the spin. This effective model applies to molecules with magnetic moment larger than 1 $\mu_B$, that is, $S>\frac{1}{2}$, and it only considers up to second order corrections in spin-orbit coupling. The sign of $D$ determines whether the system exhibits easy-axis ($D<0$) or easy-plane ($D>0$) magnetic anisotropy. 

The magnetic field in Eq.~\ref{H_spin} can be exploited in two distinct ways. First, since the GSs are twofold degenerate, they are equally likely to be occupied. For device applications, however, it is necessary to initialize the molecular system in a well-defined state. This can be achieved by applying a magnetic field along the easy axis, ($\vec{H}=H_{z}\hat{z}$). In this case, states with negative $m$ decrease in energy, whereas states with positive $m$ increase in energy. Consequently, at sufficiently low temperatures, only the $\ket{m=-S}$ state remains populated. Second, a transverse magnetic field can induce quantum tunneling through the magnetic anisotropy barrier, thereby enabling transitions between the two ground states. This mechanism is essential for switching the magnetic state of the molecule.
%

For molecules with $E=0$ (which is the case for most metallocenes), and $\vec{H}=0$, the energy levels of the system are given by $\ket{m}$, with the corresponding energies 
\begin{align}
     \epsilon_{m}=D(m^{2}-S(S+1)/3)
\end{align}
Clearly, the energy levels are pairwise degenerate, except for $m=0$ when the spin is an integer. For uniaxial molecules ($D < 0$), the degenerate groundstates (GS) are those with $|m\rangle =\pm|S\rangle$, which are central for spintronics applications, as they correspond to 0 and 1 states of the molecular device. For $\vec{H}=0$, the tunneling is prohibited between these two degenerate GSs. When a magnetic field is applied transverse to the uniaxial axis of the molecule, say in the x direction, $\vec{H}=H_x\hat{x}$, the field mixes the states differing by $\Delta m =\pm1$. In order to get the tunneling splitting $\Delta_H$ due to the in-plane magnetic field  between the lowest energy states $\ket{\pm S}$, it is required to treat the magnetic perturbation in the $2S^{th}$ order. In this case, the tunneling splitting between $\ket{\pm S}$ is given by \cite{Hartmann1995}

\begin{align}
     \Delta_{H} = \frac{4SD}{(2S-1)!} (\frac{g\mu_BH_x}{2D})^{2S}
     \label{deltaH} 
\end{align}

It is evident from Table~\ref{tab:MAE} that all metallocenes studied in this work exhibit $E=0$ with the exception of $^*$MoCp$_2$ and $^*$CrCp$_2$. As discussed in section~\ref{stability}, these molecules undergo Jahn-Teller distortion which lowers their symmetry from $S_{10}$ to $C_i$. Consequently, a finite transverse magnetic anisotropy parameter $E \neq 0$ emerges. In the case of $^*$MoCp$_2$ molecule, however, the value of E remains very small, indicating only a weak Jahn--Teller distortion. The presence of a nonzero E term enables magnetization tunneling even in the absence of an external magnetic field. The resulting tunnel-splitting $\Delta_{E,m}$ between $\ket{\pm m}$ states induced by the transverse magnetic anisotropy term (E-term in eq.~\ref{H_spin}) can be obtained from the $S^{th}$ order perturbation theory, which is given by\cite{Hartmann1995}
\begin{align}
         \Delta_{E,m} = \frac{D}{2^{3m-3}}\left(\frac{E}{D}\right)^m\frac{(S+m)!}{(S-m)![(m-1)!]^2}
  \label{deltaE}     
\end{align}
The corresponding tunneling rate of magnetization can be estimated as $\Gamma_{E,m} \approx \frac{\Delta_{E,m}^2}{ \hbar}\frac{\pi}{2}\delta(\varepsilon)$, where $\delta(\varepsilon)$ corresponds to the broadening of energy levels due to transverse anisotropy.   

To investigate tunnel-splitting in the presence of both the transverse anisotropy E and an external magnetic field $\vec{H}$, we performed an exact diagonalization of the Hamiltonian in Eq.~\ref{H_spin} using the $\ket{m}$ basis. For $^*$MoCp$_2$, for which $S=1$, tunneling is only allowed between states $\ket{1}$ and $\ket{-1}$. In contrast, for $^*$CrCp$_2$, $S=2$, and tunneling can occur between both $\ket{2} \leftrightarrow \ket{-2}$ and $\ket{1} \leftrightarrow \ket{-1}$. The zero-field tunnel-splittings obtained from the exact diagonalization are listed in Table~\ref{tab:tunnel}. The tunnel-splitting is calculated by taking the difference between the two lowest energy levels of the system. These values are in excellent agreement with those calculated using Eq.~\ref{deltaE}. We note that the tunnel-splitting associated with the $\ket{1} \leftrightarrow \ket{-1}$ transition, and consequently the corresponding tunneling rate $\Gamma_E$, is significantly larger in $^*$CrCp$_2$ than in $^*$MoCp$_2$. This enhancement arises from the substantially larger ratio (E/D) in the former molecule. By contrast, the tunnel-splitting for the $\ket{2} \leftrightarrow \ket{-2}$ transition is approximately an order of magnitude smaller, reflecting the dependence on $m=2$ in Eq.~\ref{deltaE}.
\begin{table}[ht]
\centering
\caption{The zero-field tunnel-splitting $\Delta _{E,m}$ and transition rate $\Gamma _{E,m}$ between $\ket{m}$ and $\ket{-m}$ for $^*$MoCp$_2$ and $^*$CrCp$_2$. The data are obtained from the exact diagonalization method, which is consistent with the tunnel-splitting obtained from Eq.~\ref{deltaE}.}
\vspace{\baselineskip}
\begin{tabular}{ccccccccccc} \hline\hline
\label{tab:tunnel}
Metal   & & $S$ & &  $\Delta _{E,1}$ (K)& & $\Gamma _{E,1} $ (ns$^{-1}$)& &  $\Delta _{E,2}$ (K)& & $\Gamma _{E,2} $ (ns$^{-1}$) \\ \hline
$^*$Mo & & 1  & &    0.02    & & 4.11    & & --   & & --     \\
$^*$Cr & & 2  & &    4.56    & & 937.79  & & 1.12 & & 231.24 \\ \hline\hline
\end{tabular}
\end{table}

\begin{figure}[h]
    {{\includegraphics[width=7cm]{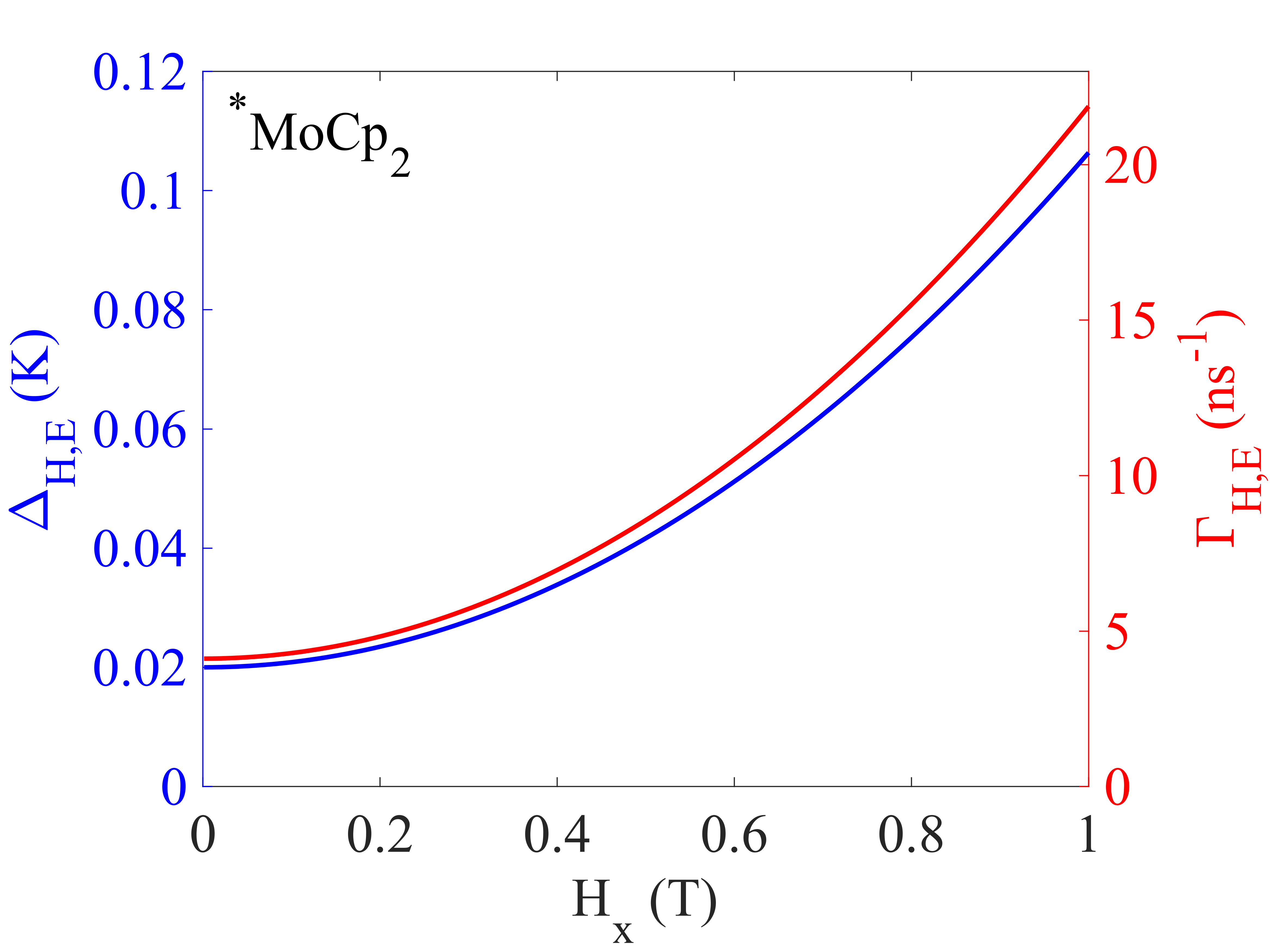}}}%
    \\
    {{\includegraphics[width=7cm]{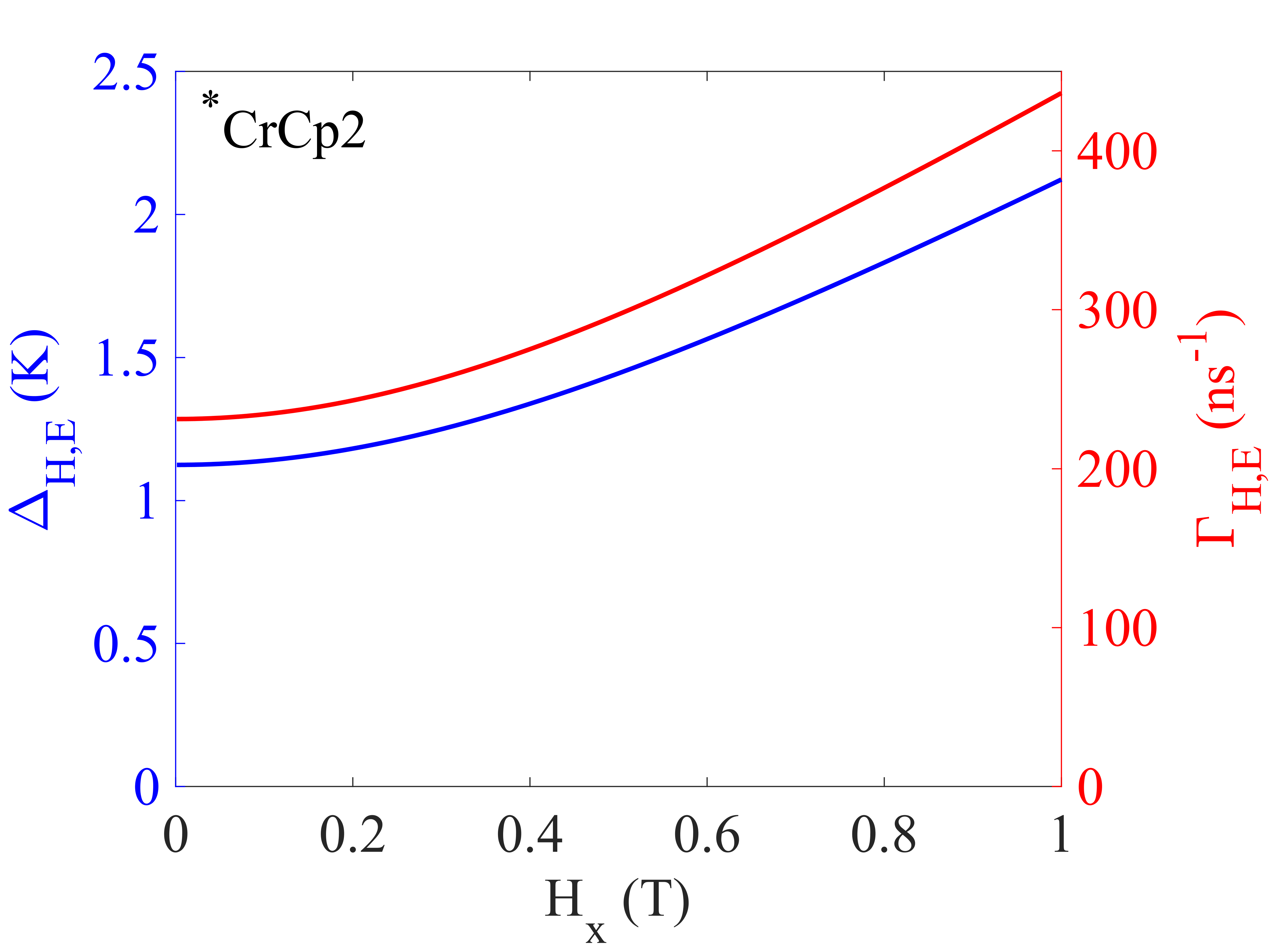}}}%
    \caption{Tunneling splitting $\Delta_{H,E}$ (blue) and transition rate $\Gamma_{H,E}$ (red) between the lowest energy states $\ket{-1}$ and $\ket{1}$ of $^*$MoCp$_2$ (top) and between the lowest energy states $\ket{-2}$ and $\ket{2}$ of $^*$CrCp$_2$ (bottom) due to both an applied transverse magnetic field $H_x$ and transverse magnetic anisotropy.}%
    \label{fig:tunnel}%
\end{figure}

The dependence of tunnel-splitting $\Delta_{H,E}$ and the corresponding tunneling rate $\Gamma_{H,E}$ between the lowest energy states on an applied in-plane magnetic field and transverse magnetic anisotropy was obtained by diagonalizing the spin Hamiltonian for different values of $H_x$. The results for $^*$MoCp$_2$ and $^*$CrCp$_2$ molecules are shown in Fig. \ref{fig:tunnel} over the field range $0 \leq H_x \leq 1 T$. A fit to the calculated tunnel-splitting for $^*$MoCp$_2$ reveals a quadratic dependence on the transverse magnetic field, $\Delta_{H,E} \propto H_x^2$. The field-induced splitting remains below 0.12 K even at $H_x = 1$ T, resulting in a relatively modest tunneling rate. In contrast, $^*$CrCp$_2$ exhibits a much stronger response to the applied field and increases at a rate $(H_x)^4$. The tunnel-splitting reaches approximately 2 K at $H_x=1$ T, while the corresponding tunneling rate approaches 440 ns$^{-1}$. This behavior can be understood from Eqs. \ref{deltaE} and \ref{deltaH}, which predicts that the tunnel-splitting induced by the transverse magnetic anisotropy scales as $E^S$, whereas the field-induced splitting scales as $H_{x}^{2S}$. Consequently, the dependence of the tunnel-splitting on the transverse field differs qualitatively for $S=1$ and $S=2$ systems. These results demonstrate that the tunneling dynamics is governed not only by the magnetic anisotropy parameters but also by the total spin of the molecule, which strongly influences the sensitivity of the tunnel splitting to external magnetic fields.

\subsection{Metallocene as magnetic memory and magnetic sensing device}
\label{application}

The electronic and magnetic properties of the $4d$ and $3d$ metallocenes investigated in this work reveal that most metallocenes exhibit easy-plane magnetic anisotropy. The notable exceptions are $^*$MoCp$_2$ and $^*$CrCp$_2$, which display uniaxial magnetic anisotropy and therefore satisfy a key requirement for magnetic memory applications. However, the large transverse anisotropy present in $^*$CrCp$_2$ gives rise to significant quantum tunneling of magnetization, enabling rapid tunneling through the anisotropy barrier, resulting in fast magnetization reversal. As a result, $^*$MoCp$_2$ emerges as the only viable candidate for magnetic memory applications among the metallocenes considered here. However, an anisotropy barrier of $\sim$ 20 K limits its usefulness only at very low temperature. 
\begin{figure}[!h]
  \includegraphics[width=0.9\linewidth]{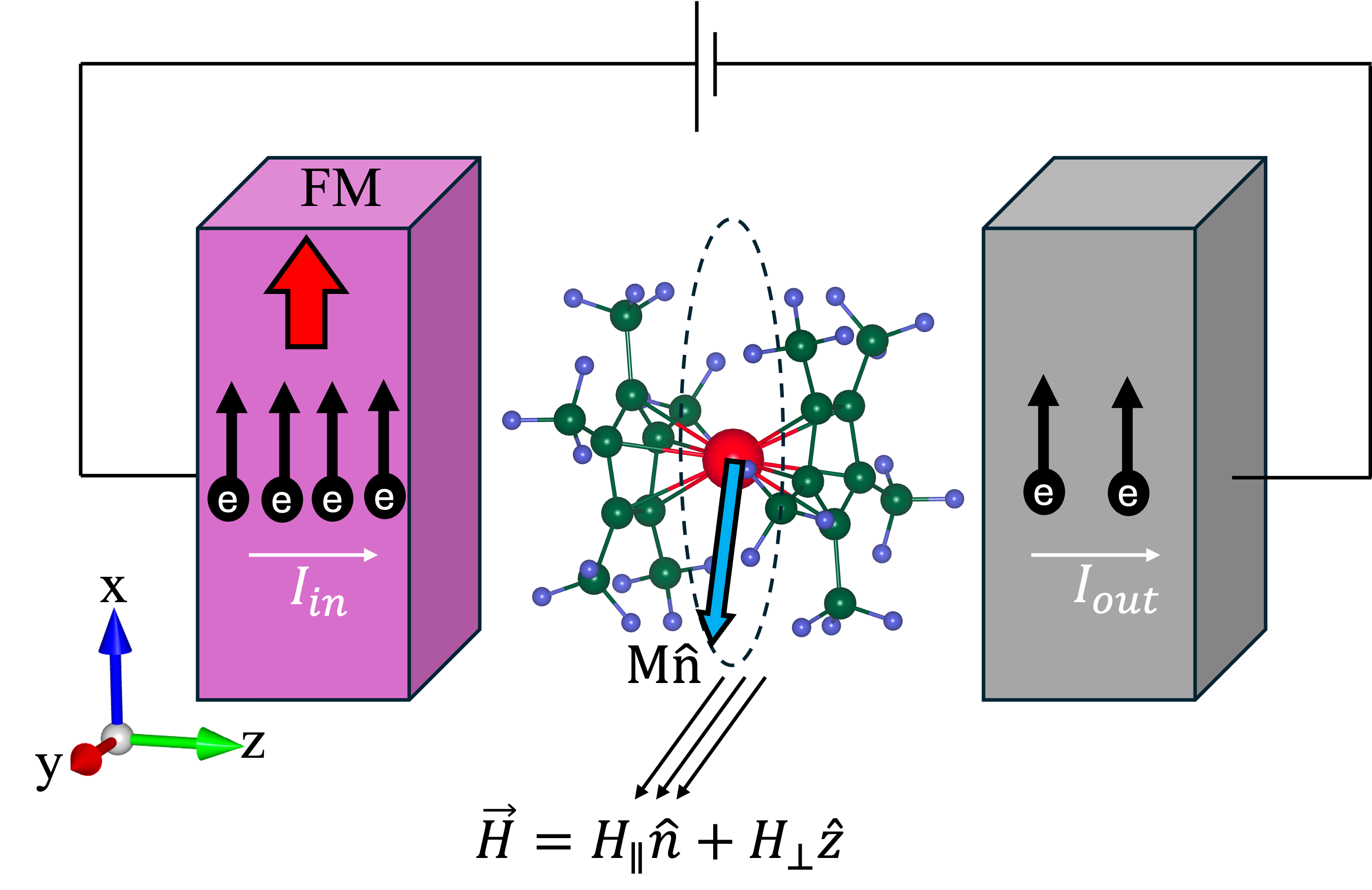}
  \caption{Schematic of proposed experimental setup to use metallocenes with easy-plane magnetic anisotropy as a weak magnetic field sensor. The left electrode is ferromagnetic. Here, $M\hat{n}$ denotes the magnetic moment of the metal center in metallocene, while $\vec{H}$ represents a weak external magnetic field with components $H_{\parallel}$, directed along the direction $\hat{n}$ in the xy plane, and $H_{\perp}$ along $\hat{z}$.}
  \label{fig:mag-sen}
\end{figure}

Although metallocenes with easy-plane anisotropy are not suitable for information storage, their magnetic properties remain of considerable interest. In particular, the strong sensitivity of their spin states to external magnetic fields may make them attractive candidates for magnetic-field sensing applications. As an illustrative example, Fig. \ref{fig:mag-sen} presents a conceptual device architecture in which a metallocene molecule is positioned between two electrodes. The left electrode consists of a ferromagnetic material (FM), producing a spin-polarized input current, $I_{in}$, while the right electrode is non-magnetic. In the absence of an external magnetic field, the magnetic moment $\vec{M}$ of the metal center can randomly orient along any direction within the xy plane since the energy barrier within the plane is negligible for metallocenes due to azymutal symmetry. Under these conditions, the spin-dependent transmission probability is $T_0 = 1/2$ (average over the xy plane), resulting in an output current, say, $I_{out}$. When a weak external magnetic field $\vec{H}$, with $H_{\parallel}$ directed along $\hat{n}$ in the xy plane and out-of-plane component $H_{\perp}$ along the molecular axis (z), the magnetic moment of metal atom in the metallocene will align along the magnetic field. This reorientation of the magnetic moment of the metal atom will change the transmission probability, which, in turn, will change the output current to $I_{H}$. The difference in electric current $\Delta I=I_{out}-I_{H}$ indicates the presence of an external magnetic field. A quantitative assessment of such a device would require explicit transport calculations, which are beyond the scope of the present work.
\vspace{\baselineskip}

\section{Summary}
\label{summary}

In this study, we have systematically investigated the structural, electronic, and magnetic anisotropic properties of metallocenes based on 4d and 3d transition metals. Our vibrational analysis reveals that most of these systems maintain structural stability within the considered ligand environment, while others experience Jahn-Teller distortions that reduce their symmetry from $S_{10}$ to $C_i$, resulting in modifications to their electronic structure. We also explore the impact of smaller ligands, observing that while ligand size influences vibrational stability, the electronic structure remains largely unaffected as long as the original $S_{10}$ symmetry is preserved. These results show that reduced-ligand models can reproduce the electronic structure of the high-symmetry configurations. However, they do not provide a reliable description of structural stability and vibrational properties.

In the metallocenes, the Cp$_2$ ligands act as electron-withdrawing groups, promoting charge transfer from the metal center to the ligands, as confirmed by our charge-density calculations. This metal–ligand hybridization significantly modifies the crystal-field splitting and the ordering of the $d$ orbitals, thus governing the magnetic properties of the molecules.

The calculated magnetic anisotropy energies indicate that these metallocenes are unlikely to possess sufficiently large uniaxial anisotropy barriers for robust magnetic memory applications. Our analysis of the 3d and 4d series reveals that the magnetic anisotropy is governed primarily by the ordering of the orbitals near the Fermi level, rather than by the number of d electrons alone. The Jahn-Teller effect lifts the orbital degeneracies of some metallocenes, which induces transverse magnetic anisotropy and therefore enables the quantum tunneling of magnetization.

\section*{Acknowledgment}

The calculations were performed on the Paro high-performance
computing system at UTEP. D. H. -M. was supported by M. F. I.'s startup grant. J.P.-M., and M.R.P. were supported by the Tec$^4$ project, funded by the U.S. Department of Energy, Office of Science, Office of Basic Energy Sciences, the Division of Chemical Sciences, Geosciences, and Biosciences (under Grant No. FWP 82037). K.P.K.W. acknowledges support from the Regents’ Research Excellence Program of The University of Texas System at The University of Texas at El Paso.

\bibliography{metallocene.bib}

\end{document}